\documentclass[10pt,journal,compsoc]{IEEEtran}

\usepackage{dblfloatfix} 

\input{dependencies}

    \definecolor{ao(english)}{rgb}{0.0, 0.5, 0.0}
    
	\definecolor{armygreen}{rgb}{0.29, 0.33, 0.13}
\begin{document}
\title{Finding   Trends in Software  Research}

\author{George~Mathew,
        Amritanshu~Agrawal 
        and Tim~Menzies,~\IEEEmembership{Senior Member~IEEE}%
        
\IEEEcompsocitemizethanks{\IEEEcompsocthanksitem George Mathew is a Ph.D candidate in the Department
of Computer Science(CS) at North Carolina State University(NCSU)\protect\\
E-mail: george.meg91@gmail.com
\IEEEcompsocthanksitem Amritanshu Agrawal is a Ph.D candidate in the Department
of CS at NCSU\protect\\
E-mail: aagrawa8@ncsu.edu
\IEEEcompsocthanksitem Tim Menzies is a full professor in the Department
of CS at NCSU\protect\\
E-mail: timm@ieee.org}
}

\markboth{IEEE Transactions in Software Engineering, ~Vol.~XXX, No.~XX, August~XXXX}%
{Mathew \MakeLowercase{\textit{\etal}}: On the Value of Text Mining for Detecting Trends in Software Engineering Research.}

\IEEEtitleabstractindextext{%
\begin{abstract}

This paper explores the structure of research papers in software engineering. Using text mining, we study 35,391 software  engineering (SE) papers from 34 leading SE venues over the last 25 years. These venues were divided, nearly evenly, between conferences and journals.  An important aspect of this analysis is that it is fully automated and repeatable.   To achieve that automation, we used   topic modeling (with LDA) to mine   10 topics that represent much of the structure of contemporary SE.  The  10 topics presented here should not be ``set in stone'' as the only  topics   worthy of study in SE.  Rather our goal is to report that   (a)~text mining methods can  detect large scale trends within our community;  (b)~those topic change with time; so (c)~it is  important to have automatic agents that can update our understanding of our community whenever new data arrives.  

\end{abstract}
}

\maketitle

\IEEEdisplaynontitleabstractindextext

\IEEEpeerreviewmaketitle

\ifCLASSOPTIONcompsoc
\IEEEraisesectionheading{\section{Introduction}\label{sect:intro}}
\else
\section{Introduction}
\label{sect:intro}
\fi

Each year, SE researchers spend much effort writing and presenting papers to   conferences, and journals.
What can we learn about all those papers? What are the
factors that prevent effective dissemination of research results?
If we study patterns of acceptance in our SE papers,
can  we improve how we do, and report, research in software engineering?

Such  introspection lets us answer important questions like: 
\be
    \item {\em Hot topics:} What is  
    the hottest topic in SE research? How is this list of ``hot topics''
    changes over time?
     \item {\em Breadth vs depth?:} Should a researcher focus on one particular research topic or venture across multiple topics?
    \item {\em Gender Bias:} Is there a gender based bias in SE research?
   \item {\em Where to publish:  journals or conference?:}  What would be the ideal venue for your latest research?
\ee
To answer such questions as these,
this paper applies  text mining and clustering
using LDA (Latent  Dirichlet  Allocation) to 
 35,391 SE papers from the last
25 years published in 34 top-ranked conferences and journals.
These   venues are then clustered by   what topics they share.
Using these information, we have answers to many questions, including those shown above.
Using these results,
 journal and conference organizers   could find more relevant reviewers for a particular paper much faster.
   Also, research program managers could find more relevant reviewers much faster
    for grant proposals.
    Lastly, the administrators of funding bodies could determine which
    areas of research are over/under-represented and, hence, those that are worthy of less/more funding (respectively).

In summary, this paper makes the following contributions:
\bi
\item
A fully automatic and repeatable process for discovering a topology between tens of thousands
of technical papers. 
\item An open-source toolkit that other researchers can apply to other sets of documents: see \href{https://goo.gl/Au1i53}{goo.gl/Au1i53}.
\item A large database that integrates all our results into one, easy-to-query SQL database: see \href{https://goo.gl/zSgb4i}{goo.gl/zSgb4i}.
\item A 
lightweight approach to gathering information about the entire SE community. Using
that information, we can not
only answer the four questions above, but   many more like it. 
\ei
Further to the last point,
the answers to our four questions are:


\begin{table}[!t]
{\footnotesize
  \begin{center}
\begin{tabular}{r@{~=~}p{2.5in}}\hline
  Source Code & code, source, information, tool, program, developers, patterns \\
  {Software process} & requirements, design, systems, architecture,\newline analysis, process, development \\
  Modeling & model, language, specification, systems, techniques, object, uml \\
  Program Analysis & program, analysis, dynamic, execution, code, java, static \\
  Metrics & metrics, data, quality, effort, prediction, defect, analysis \\
  Developer & developer, project, bug, work, open, team, tools \\
  Applications & applications, web, systems,  component,\newline  services, distributed, user \\
  Testing & test, testing, cases, fault, techniques, coverage, generation \\
  Performance & performance, time, data, algorithm, systems, \newline  problem, network, distributed \\
  Architecture & architecture, component, systems, design, product, \newline reuse, evolution  
   \\\hline

\end{tabular}
\end{center}}
\caption{The top 7 terms in the 10 major SE topics
as found by this research.   Topics are ordered top-to-bottom, most-to-least frequent. Also, terms within topics are ordered left-to-right most-to-least frequent. Topics are 
named using the most frequent terms. }
\label{tab:11topics}
\end{table}

\begin{table*}
\fontsize{8}{10}\selectfont
\centering
\begin{tabular}{|c|r|l|c|c|c|c|}
\hline
\textbf{Index} & \multicolumn{1}{|l|}{\textbf{Short}} & \multicolumn{1}{c|}{\textbf{Name}} & \textbf{Type} & \textbf{Start}  & \textbf{h5} & \textbf{Group}\\ \hline

1 & MDLS & International Conference On Model Driven Engineering Languages And Systems & Conf & 2005 & 25 &  A1\\ 
& SOSYM & Software and System Modeling & Jour & 2002 & 28 & A2\\\hline

2 & S/W & IEEE Software & Jour & 1991 & 34 & B1\\\hline

3 & RE & IEEE International Requirements Engineering Conference & Conf & 1993 & 20 &  C1\\ 
& REJ & Requirements Engineering Journal & Jour & 1996 & 22 & C2\\\hline

4 & ESEM & International Symposium on Empirical Software Engineering and Measurement & Conf & 2007 & 22 & D1 \\
& ESE & Empirical Software Engineering & Jour & 1996 & 32 & D2\\ \hline

5 & SMR & Journal of Software: Evolution and Process & Jour & 1991 & 19 & E1\\
& SQJ & Software Quality Journal & Jour & 1995 & 24 & E2\\ 
& IST & Information and Software Technology & Jour & 1992 & 44 & E3\\ \hline

6 & ISSE & Innovations in Systems and Software Engineering & Jour & 2005 & 12 & F1\\
& IJSEKE & International Journal of Software Engineering and Knowledge Engineering & Jour & 1991 & 13 & F2\\
& NOTES & ACM SIGSOFT Software Engineering Notes & Jour & 1999 & 21 & F3\\ \hline

7 & SSBSE & International Symposium on Search Based Software Engineering & Conf & 2011 & 15 & G1\\
& JSS & The Journal of Systems and Software & Jour & 1991 & 53 & G2\\
& SPE & Software: Practice and Experience & Jour & 1991 & 28 & G3\\ \hline

8 & MSR & Working Conference on Mining Software Repositories & Conf & 2004 & 34 & H1\\
& WCRE & Working Conference on Reverse Engineering & Conf & 1995 & 22 & H2\\
& ICPC & IEEE International Conference on Program Comprehension & Conf & 1997 & 23 & H3\\
& ICSM & IEEE International Conference on Software Maintenance & Conf & 1994 & 27 & H4\\
& CSMR & European Conference on Software Maintenance and Re-engineering & Conf & 1997 & 25 & H5\\
 \hline

9 & ISSTA & International Symposium on Software Testing and Analysis & Conf & 1989 & 31 & I1\\
& ICST & IEEE International Conference on Software Testing, Verification and Validation & Conf & 2008 & 16 & I2\\
& STVR & Software Testing, Verification and Reliability & Jour & 1992 & 19 & I3\\ \hline

10 & ICSE & International Conference on Software Engineering & Conf & 1994 & 63 & J1\\
& SANER & IEEE International Conference on Software Analysis, Evolution and Re-engineering & Conf & 2014 & 25 & J2\\ \hline

11 & FSE & ACM SIGSOFT Symposium on the Foundations of Software Engineering & Conf & 1993& 41 & K1\\
& ASE & IEEE/ACM International Conference on Automated Software Engineering & Conf & 1994 & 31 & K2\\ \hline

12 & ASEJ & Automated Software Engineering Journal & Jour & 1994 & 33 & L1\\
& TSE & IEEE Transactions on Software Engineering & Jour & 1991 & 52 & L2\\
& TOSEM & Transactions on Software Engineering and Methodology & Jour & 1992 & 28 & L3\\ \hline

13 & SCAM & International Working Conference on Source Code Analysis \& Manipulation & Conf & 2001 & 12 & M1\\
& GPCE & Generative Programming and Component Engineering & Conf & 2000 & 24 & M2\\
& FASE & International Conference on Fundamental Approaches to Software Engineering & Conf & 1998 & 23 & M3\\ 
 \hline

\end{tabular}
\caption{Corpus of venues (conferences and journals) studied in this paper. 
For a rationale of why these venues were selected, see \tion{data}.
Note two recent changes to the above names:
ICSM is now called ICMSE; and WCRE and CSMR recently fused into SANER.
In this figure,
the ``Group'' column shows venues that publish ``very similar'' topics (where similarity is computed via a cluster analysis shown later in this paper). The venues are selected in a diverse range of their h5 scores between 2010 and 2015. h5 is the h-index for articles published in a period of 5 complete years obtained from Google Scholar. It is the largest number ``h'' such that ``h'' articles published in a time period have at least ``h'' citations each.}
\label{tab:venues}
\end{table*} 

\begin{table*}

\fontsize{8}{10}\selectfont
\begin{tabular}{|l|p{16cm}@{~}|}\hline
\textbf{Topic} &\textbf{Top Papers} \\ \hline
 Program Analysis &
 1992: Using program slicing in software maintenance; KB Gallagher, JR Lyle \vspace{-0.5mm}\newline
 2012: Genprog: A generic method for automatic software repair; C Le Goues, TV Nguyen, S Forrest, W Weimer
 \\ \hline
\textcolor{black}{ Software process} &
1992: A Software Risk Management Principles and Practices; BW Boehm \vspace{-0.5mm}\newline
2009: Seven Process Modeling Guidelines (7PMG); J Mendling, HA Reijers, WMP van der Aalst
\\\hline
 Metrics &
1996: A validation of object-oriented design metrics as quality indicators; VR Basili, LC Briand, WL Melo \vspace{-0.5mm}\newline
2012: A systematic literature review on fault prediction performance in software engineering; T Hall, S Beecham, D Bowes, D Gray, S Counsell
\\\hline
 Applications &
  2004: Qos-aware middleware for web services composition;
L Zeng, B Benatallah, AHH Ngu, M Dumas,  
J Kalagnanam, H Chang\vspace{-0.5mm}\newline
  2011: CloudSim: a toolkit for modeling \& simulation of cloud computing; R.Calheiros, R.Ranjan, A.Beloglazov, C.De Rose, R.Buyya
  \\\hline
 Performance &
1992: Spawn: a distributed computational economy; CA Waldspurger, T Hogg, BA Huberman \vspace{-0.5mm}\newline 
2010: A theoretical and empirical study of search-based testing: Local, global, and hybrid search; M Harman, P McMinn
\\ \hline
Testing &
 2004: Search-based software test data generation: A survey;
P McMinn \vspace{-0.5mm}\newline
 2011: An analysis and survey of the development of mutation testing; Y Jia, M Harman  \\ \hline
Source Code &
 2002: CCFinder: A Multilinguistic Token-Based
Code Clone Detection System
for Large Scale Source Code; T Kamiya, S Kusumoto, K Inoue \vspace{-0.5mm}
 2010: DECOR: A method for the specification and detection of code and design smells; N Moha, YG Gueheneuc, L Duchien, AF Le Meur\\ \hline
Architecture &
2000: A Classification and Comparison Framework for Software Architecture Description Languages; N Medvidovic, RN Taylor \vspace{-0.5mm}\newline
2009: Software architecture many faces many places yet a central discipline; RN Taylor
\\\hline
 Modelling &
 1997: The model checker SPIN; GJ Holzmann \vspace{-0.5mm}\newline
 2009: The ``physics" of notations: toward a scientific basis for constructing visual notations in software engineering; D Moody \\ \hline
Developer & 
2002: Two case studies of open source software development Apache and Mozilla; A Mockus, RT Fielding, JD Herbsleb\vspace{-0.5mm}\newline
 2009: Guidelines for conducting and reporting case study research in software engineering; P Runeson, M Host 
 \\ \hline
\end{tabular}
\caption{ 
Most cited papers within our 10 SE topics. The first paper in each row is the top paper  since 1991 and the second paper is the top paper for the topic since 2009.
For a definition of these topics, see \tref{11topics}.}
\label{tab:topPapersMixed}
\end{table*}

\be
\item Regarding {\em hot topics}: We  identify 10 major topics currently in software engineering; see \tref{11topics}.
Also, we show  how those groupings have changed over recent years. 
\item The creation of a ``reader'' of the top venues in SE (see \tref{venues})
that  lists the   most-cited papers in our topics  (see \tref{topPapersMixed}).
\item Regarding {\em breadth vs depth:} We find that mono-focusing on a single topic can lead
to fewer citations than otherwise.
\item Regarding {\em gender bias: }  We have mixed news here. The percentage of women researchers  in SE (22\%) is much larger than other fields (i.e., 11\% in mathematics and 13\% in economics). Also, of the top 1000 most cited SE authors (out of 35,406),  the percentage of published women
is on par with the overall percentage of women in this field. That said,
of the top 10 most cited authors, only one is female.   
\item Regarding {\em where to publish}, we offer a previously unreported dichotomy between software conferences and journals. As shown in \tion{study_conf_vs_jour},
SE conference publications tend to publish on different topics to SE journals
and those conferences publications earn a significantly larger number of citations than journal articles (particularly in the last four years).
\ee
The rest of this paper is structured as follows.   We start of with the description of the data used in this study from various sources and its consolidation in \tion{data}. This is followed by a description of how we used   topic modeling     to find  representative topics in SE.  
 Next, \tion{sanity} finds that the  topics generated by topic modelling are reasonable.
Hence, \tion{results} goes on to highlight important case studies which can be used using the proposed method.  The threats to the validity of this study is described in  \tion{threats}. \tion{rel_work} gives an overview of prior and contemporary studies based on bibliometrics and topic modeling in SE literature. \tion{conclusion} concludes this study by presenting the inferences of this study and how it can aid the different sections of SE research community.


Note that some of the content of this paper was presented as a  short two-page paper previously published in the ICSE-17 companion~\cite{Mathew2017}. Due to its small size,
that document discussed very little of the details discussed here.

\section{Data}
\label{sect:data}

For studying the trends of SE, we build a repository of 35,391 papers and 35,406 authors from 34 SE venues over a period of 25 years between 1992-2016. This time period (25 years) was chosen since it encompasses recent trends in software engineering
such as the switch from waterfall to agile; platform migration from desktops to mobile; and the rise of cloud computing. Another reason to select this 25 year cut off was that we 
encountered increasingly more difficulty in accessing data prior to 1992; i.e., before the widespread use of the world-wide-web.

As to the venues used in this study, these were selected via a combination of on-line citation indexes (Google Scholar), feedback from the international research community (see below) and our own domain
expertise:
\bi
\item
Initially, we selected all the non-programming language peer-reviewed conferences
from the ``top publication venues'' list of Google Scholar Metrics (Engineering and Computer Science, Software Systems).  Note that
Google Scholar generates this list based on citation counts.
\item
Initial feedback from conference  reviewers (on a rejected earlier version
of this paper) made us look also at SE journals.
\item
To that list, using our domain knowledge, we added
venues that we knew were associated with senior researchers in the field;
e.g., the ESEM and SSBSE conferences. 
\item
Subsequent feedback from
an ICSE'17 companion presentation~\cite{Mathew2017} about this
work made us add in journals and conferences related to modeling. 
\ei
This resulted in the venue list of \tref{venues}.   


For studying and analyzing those venues we construct a database of 18 conferences, 16 journals, the papers published with the metadata, authors co-authoring the papers and the citation counts from 1992-2016. Topics for SE are generated using the titles and abstracts of the papers published in these venues rather than the entire text of the paper. Titles and abstracts have been widely used in text based bibliometric studies~\cite{sarkar2017predicting, garousi2016citations, cai2008analysis} primarily due to three reasons: (a) Titles and abstracts are designed to index and summarize papers; (b) Obtaining papers is a huge challenge due to copyright violations and its limited open source access; (c) Papers contain too much text which makes it harder to summarize the content. Abstracts on the other hand are much more succinct and generate better topics.

The data was collected in five stages:
\be
\item Venues are first selected manually based on top h5-index scores from Google Scholar. It should be noted that all the data collection method is automated if the source of papers from a desired venue is provided. Thus, this can be expanded to additional venues in the future.
\item For each venue,  DOI (Document Object Identifier), authors, title, venue \& year for every publication between 1992-2016 is obtained by scrapping the html page of DBLP\footnote{\url{http://dblp.uni-trier.de/}}. DBLP (\textbf{D}ata\textbf{B}ase systems \& \textbf{L}ogic \textbf{P}rogramming) computer science bibliography is an on-line reference for bibliographic information on major computer science publications. As of Jan 2017, dblp indexes over 3.4 million publications, published by more than 1.8 million authors. 
\item For each publication, the abstract is obtained from the ACM portal via AMiner\footnote{\url{https://aminer.org/citation}}~\cite{Tang:08KDD} periodically on their website. Abstracts for 21,361 papers from DBLP can be obtained from the ACM dump. For rest of the papers, we use only the tiles for the subsequent study.
\item For each publication, we obtain the corresponding citation counts using crossref's\footnote{\url{https://www.crossref.org/}} REST API.
\item The final stage involves acquiring the gender of each author. We used the opensource tool\footnote{\url{https://git.io/vdtLp}} developed by Vasilescu \etal in their quantitative study of  gender representation and online participation~\cite{vasilescu2013gender}. We used a secondary level of resolution for names that the tool could not resolve by referring the American census data\footnote{\url{https://www.census.gov/2010census/}}. Of the 35,406 authors in SE in our corpus we were able to resolve the gender for 31,997 authors.
\ee
Since the data is acquired from different sources, a great challenge lies in merging the documents. There are two major sets of merges in this data:
\bi
\item \textit{Abstracts to Papers}: A merge between a record in the ACM dump and a record in the DBLP dump is performed by comparing the title, authors, venue and the published year. If these four entries match, we update the abstract of the paper in the DBLP dump from the ACM dump. To verify this merge, we inspected 100 random records manually and found all these merges were accurate.
\item \textit{Citation Counts to Papers}: DOIs for each article can be obtained from the DBLP dump, then   used to query crossref's rest API to obtain  an approximate of the citation count. Of the 35,391 articles, citation counts were retrieved for 34,015 of them.
\ei

\section{Topic Modeling}
\label{sect:lda}

\begin{wrapfigure}{r}{1in}
  \centering
  \captionsetup{justification=centering}
  \includegraphics[width=1in]{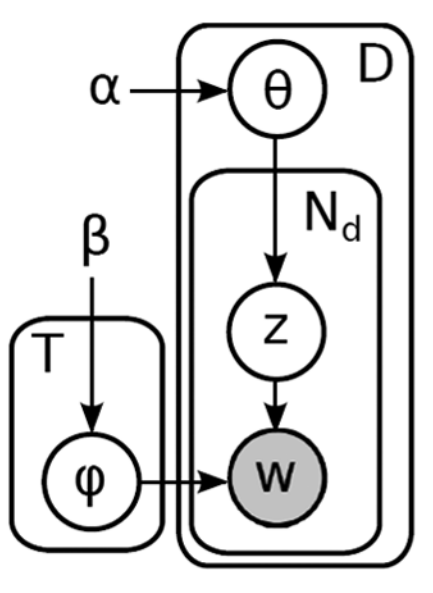}
  \caption{LDA}\label{fig:lda}
\end{wrapfigure}
 
\begin{figure*}[!b]
\centering
\includegraphics[scale=0.75]{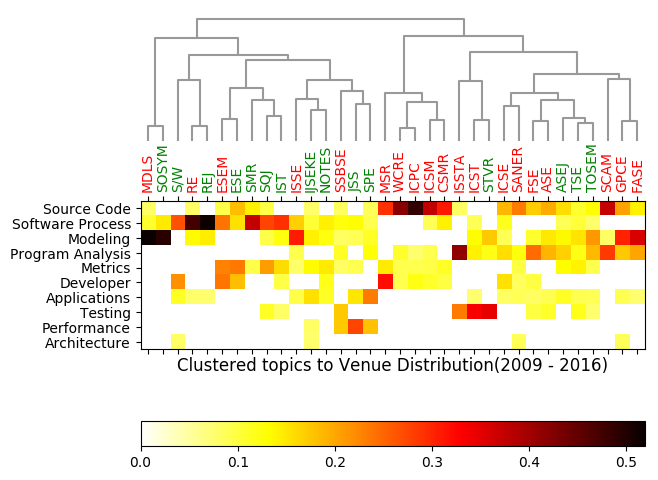}
\caption{Hierarchical clustering heatmap of Topics and Venues between the years 2009-2016. 
Along the top,  \textcolor{ao(english)}{{\bf green}} denotes
journals and   \textcolor{red}{{\bf red}} denotes a conference.
Along the left side, the black words are the  topics of \tref{11topics}. 
Topics are sorted top-to-bottom, most-to-least frequent.
Each cell in the heatmap depicts the 
frequency of a topic in a particular
venue.  The tree diagrams above the venues 
show the results of a bottom up clustering of the venues with respect to the topics. In those clusters, venues are in similar
sub-trees if they have similar pattern of topic frequencies.}
\label{fig:div_heat}
\end{figure*}

As shown in Figure~\ref{fig:lda},
the LDA topic modelling algorithm~\cite{blei2003latent,
nikolenko2015topic} assumes   
  $D$ documents
 contain \textit{T} topics expressed with \textit{W}
different words.  Each document $d \in D$ of length $N_d$ is modeled
as a discrete distribution $\theta(d)$ over the set of topics. Each
topic corresponds to a multinomial distribution over the
words. $\alpha$ is the discrete prior assigned to the distribution of
topics vectors($\theta$); and $\beta$ for the distributions of words in
topics($\psi$).

As shown in Figure~\ref{fig:lda}, the outer plate spans documents and the inner
plate spans word instances in each document (so the \textit{w} node
denotes the observed word at the instance and the \textit{z} node
denotes its topic). The inference problem in LDA is to find hidden
topic variables \textit{z}, a vector spanning all instances of all
words in the dataset.  LDA is a problem of Bayesian inference. The
original method used is a variational Bayes approximation of the
posterior distribution~\cite{blei2003latent} and alternative inference
techniques use Gibbs sampling~\cite{griffiths2004finding} and
expectation propagation~\cite{minka2002expectation}.

There are many examples of the use of LDA in SE.
For example, Rocco et al.~\cite{oliveto2010equivalence} used  text mining and Latent Dirichlet Allocation
(LDA) for traceability link recovery.  Guzman and Maleej perform sentiment analysis on App Store reviews to identify fine-grained app features~\cite{guzman2014users}. The features are identified using LDA and augmented with sentiment analysis. Thomas \etal use topic modeling in prioritizing static test cases~\cite{thomas2014static}.

Topic modeling is powered by three parameters; 1) $k$: Number of topics 2) $\alpha$ : Dirichlet prior on the per-document topic distributions 3) $\eta$: Dirichlet prior on the per-topic word distribution.

To set these parameters, we   used {\em perplexity} to find the best $k$ number of topics.
Perplexity is the probability of all the words in an untrained document given the topic model, so {\em most} perplexity is {\em best}.
To use perplexity, we varied topics from 2 to 50, then ran  a 20 fold cross-validation study (95\%/5\% 
splits of data, clusters generated on train, perplexity assessed on test). Those runs  found no significant   improvement
in perplexity after $k=10$.

Next, we had to find values for $\{\alpha,\eta\}$.
By default, LDA sets these to 
$\alpha = \frac{1}{\#topics} = 0.05$ and $\eta = 0.01$.
 The simplest way of tuning them would be to explore the search space of these parameters exhaustively. This approach is called grid-search and was used in tuning topic modeling ~\cite{asuncion2009smoothing}. The prime drawback of this approach is the time complexity and can be very slow for tuning real numbers. Alternatively,  numerous researchers recommend tuning these parameters using Genetic Algorithms(GA) ~\cite{panichella2013effectively, lohar2013improving, sun2015msr4sm}. In a GA, a population of candidate solutions are mutated and recombined towards better solutions. GA is also slow when the candidate space is very large and this is true for LDA since the candidates ($\alpha$ and $\beta$) are real valued.
 Hence, inspired from the work by Fu \etal on tuning learners in SE~\cite{fu2016tuning},
 we use  Differential Evolution (DE) for parameter tuning.   
Differential evolution  randomly picks three different vectors  
$B,C,D$ from a list called $F$ (the {\em frontier}) for each parent vector A in $F$ ~\cite{storn1997differential}. 
Each pick generates a new
vector $E$ (which replaces $A$ if  it scores better).
$E$ is generated as follows:
\begin{equation} \label{eq:de}
  \forall i \in A,  E_i=
    \begin{cases}
      B_i + f*(C_i - D_i)& \mathit{if}\;  \mathcal{R} < \mathit{cr}  \\
      A_i&   \mathit{otherwise}\\ 
    \end{cases}
\end{equation}
where $0 \le \mathcal{R} \le 1$ is a random number,
and $f,cr$ are constants that represent mutation factor and crossover factor respectively (following
Storn et al.~\cite{storn1997differential}, we use $cr=0.3$ and $f=0.7$).
Also, one value from $A_i$  (picked at
random)
is moved to $E_i$ to ensure that $E$ has at least one
unchanged part of an existing vector. The objective of DE is to maximize the Raw Score ($\Re_n$) which is similar to the Jaccard Similarity~\cite{galvis2013analysis}. Agrawal \etal~\cite{agrawal2016wrong} have explained the $\Re_n$ in much more detail. 
After tuning on the data, we found that optimal values in this domain are $k$, $\alpha$ \& $\beta$ are 11, 0.847 \& 0.764 respectively.

\section{Sanity Checks}
\label{sect:sanity}

Before delving into the results of the research it is necessary to critically evaluate the topics generated by LDA and the clustering of venues.

\subsection{Are our Topics Correct?}
\label{sect:clusters}
   
We turn to \fref{div_heat} to check the sanity of the topics. \fref{div_heat} shows  the results of this hierarchical clustering technique on papers published between 2009 and 2016. Topics are represented on the vertical axis and venues are represented on the horizontal axis. The \textcolor{ao(english)}{{\bf green}} venues represents   journals and the \textcolor{red}{{\bf red}} represents  conferences. Each cell in the heatmap indicates the contribution of a topic in a venue. Darker colored cells indicate strong contribution of the topic towards the venue while a lighter color signifies a weaker contribution. Venues are clustered with respect to the distribution of topics using the linkage based hierarchical clustering algorithm on the {\em complete} scheme~\cite{mullner2011modern}. The linkage scheme determines the distance between sets of observations as a function of the pairwise distances between observations. In this case we used the euclidean distance to compare the clusters with respect to the distributions of topics in each venue in a cluster. In a complete linkage scheme the maximum distance between two venues between clusters is used to group the clusters under one parent cluster.
\[max\{euclid(a,b) : a \in cluster_A, \ b \in cluster_B\}\]
The clusters   can be seen along the vertical axis of \fref{div_heat}. Lower the height of the dendogram, stronger the cohesion of the clusters.
Several aspects of  \fref{div_heat} suggest that our topics are ``sane''.
Firstly, in that figure, we can see several examples of specialist conferences paired with the appropriate journal:
\bi
\item The main modeling conference and journal (MDLS and SOSYM)
are grouped together;
\item The requirements engineering conference and journal are grouped together: see RE+REJ;
\item
The empirical software engineering conference and journals are also grouped: see ESEM+ESE;
\item Testing and verification venues 
are paired; see ICST+STVR.
\ei
Secondly, the topics learned by LDA occur at the right frequencies
 in the right venues:
 \bi
 \item The modeling topics appears most frequently in the modeling
 conference and journal (MDLAS and SOSYM);
 \item The {software process} topic appears most frequently in RE and REJ;
 i.e., the requirements engineering conference and journal.
 \item
 The testing topic appears most frequently in the venues devoted to testing: ISSTA,   ICST, STVR;
 \item
 The metrics topic occurs most often at venues  empirically assess software engineering methods: ESEM and ESE.
 \item
 The source code topic occurs most often at venues that focus most
 on automatic code analysis:  ICSM, CSMER, MSR, WCRE, ICPC.
 \item Generalist top venues like ICSE, FSE, TSE and TOSEM have no outstanding bias towards any particular topic. This is a useful
 result since, otherwise, that would
 mean that supposedly ``generalist venues'' are actually blocking the publication of certain kinds of papers.
 \ei
Thirdly, when we examine the most-cited papers that fall into our topics, it can be seen that these papers nearly always clearly correspond to our topics names 
(exception: the {``software process''} topic, discussed below).

\subsection{Are 10 Topics Enough? }
\label{sect:topicSanity}

 After instrumenting the internals of  LDA, we can report that the 
 10 topics in \tref{11topics}  covers over 95\% of the papers. While increasing the number of topics post 10 reported no large change in perplexity, those occur at diminishingly low frequencies. As evidence of this, recall that the rows of  \fref{div_heat} are sorted
 top-to-bottom most-to-least frequent. Note that the bottom few
 rows are mostly white (i.e., occur at very low frequency) while
 the upper rows are much darker (i.e., occur at much higher 
 frequency). That is, if we reported {\em more} than 10 topics
 then the 11th, 12th etc would occur at very low frequencies. Thus all the less frequent topics are grouped into a single topic called Miscellaneous.
 
  We note that this study is not the only one to conclude that
   SE can be reduced to 10 topics
Other researchers~\cite{datta2016long,garousi2013bibliometric,cai2008analysis} also report that 90\% of the topics can be approximated by about a dozen topics.

 \subsection{Are Our Topics Correctly Labelled?}
 \label{sect:topicLabel}

 Another question to ask is whether or not the topics
 of \tref{11topics} have the correct labels.
 For example, we have assigned the label ``Program analyis'' to the list of terms ``program, analysis, dynamic, execution, code, java, static''. More generally, we have labelled all our topics using first one or two words
 (exception: {``software process''}, which is discussed below). Is that valid? 
 
  \begin{figure}[!t]
\centering
\includegraphics[scale=0.5]{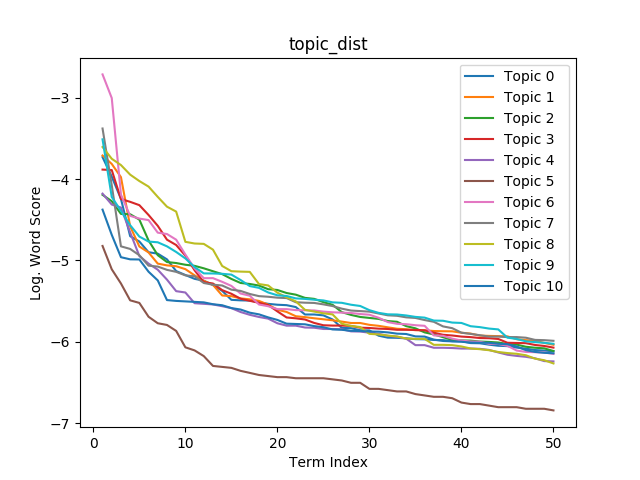}
\caption{Log Score of terms in each topic.}
\label{fig:topic_dist}
\end{figure}

 We can verify this question two different ways:
 \bi
    \item \textit{Mathematically}: \fref{topic_dist} shows  the LDA score for each term in our topics. The x-axis orders the terms in same order as the right-hand-column of \tref{11topics}. The y-axis of that plot logarithmic; i.e., there is very little information associated with the latter words. It can be seen that terms from one topic(Miscellaneous) has relatively lower scores compared to other topics. This further justifies the grouping of all the lesser topics into a single topic.
    \item \textit{Empirically}: \tref{topPapersMixed} enlists the top papers associated with each topics. The abstracts of these papers contain the terms constituting the topics.
 \ei
Overall, the evidence to hand suggests that generated labels from the first few terms is valid.
  That said, one topic was particularly challenging to label.
   For papers in the {``software process''} topic, we found a wide variety of research directions. For example, the title
 of two papers listed in this topic in  Table~\ref{tab:topPapersMixed}
 includes ``software risk management'' and ``process modeling guidelines''. Other top-cited papers in this group discussed methods for configuration and dynamic analysis of software tools.
As we read more papers from this topic, a common theme emerged; i.e. how to decide 
 between possible  parts from a set of risk, process, or product options.

\subsection{Do topics adapt to change in venues?}
\label{sect:topicAdapt}

\begin{figure}[!t]
\centering
\includegraphics[scale=0.45]{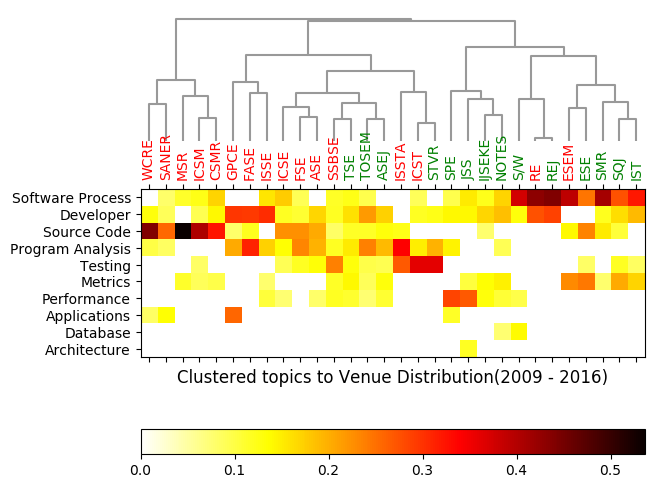}
\caption{Are our topics stable?
Here, we show a heat map of topics in venues
(2009-2016) after removing four
$MDLS, SOSYM, SCAM, ICPC$. Note
that even with those removals,
we find nearly the same patterns as
 \fref{div_heat}.}
\label{fig:div_heat_red}
\end{figure}

Finally we check how the topics vary when venues are added. It should be noted that no study can cover all papers from all venues.
Hence, it is important to understand how the addition of
more venues might change the observed structures.  For our work,
we kept adding new venues until our topics stabilizes.
To demonstrate that stability,  we show here  
the structures seen  {\em before} and {\em after} removing the
last 
venues added to this study: 
\bi
\item $MDLS$ and $SOSYM$: These two venues focused extensively on modeling. It should be noted that modeling was the third most frequent topic in the literature during 2009-2016 and was primarily due to these two venues.
\item $SCAM$: A venue which heavily focuses on Source Code and Program Analysis.
\item $ICPC$: Another venue that focuses heavily on Source Code.
\ei
Once these venues were removed, we performed LDA using the same hyper-parameters. The resultant clustered heatmap is shown in \fref{div_heat_red}. We can observe that
\be
\item The topic ``Modeling'' is now not present in the set of topics anymore. Rather a weak cluster called ``Database" is formed with the remnants.
\item {``Software process''} is now the most popular topic since none of the venues removed were contributing much towards this topic.
\item ``Source Code" has gone down the popularity since two venues contributing towards it were removed.
\item ``Program Analysis" stays put in the rankings but ``Developer" goes above it since one venue contributing towards ``Program Analysis" was removed.
\item The sanity clusters we identified in \tion{topicSanity} stays intact implying that the clustering technique is still stable.
\ee
Based on these results, we can see that a) on large changes to venues contributing towards topics, the new topics are formed and the existing topics almost stays intact; b) on small changes to venues contributing towards topics, the topics either become more or less prominent but new topics are not created. Thus, we can see that the topics produced by this approach is relatively adaptable.

\ifCLASSOPTIONcaptionsoff
  \newpage
\fi

\section{Discussion}
\label{sect:results}
\subsection{What are the Hottest topic in SE Research?}
\label{sect:study_hot}

\begin{figure}[!htpb]
\centering


{\centering \includegraphics[scale=0.45]{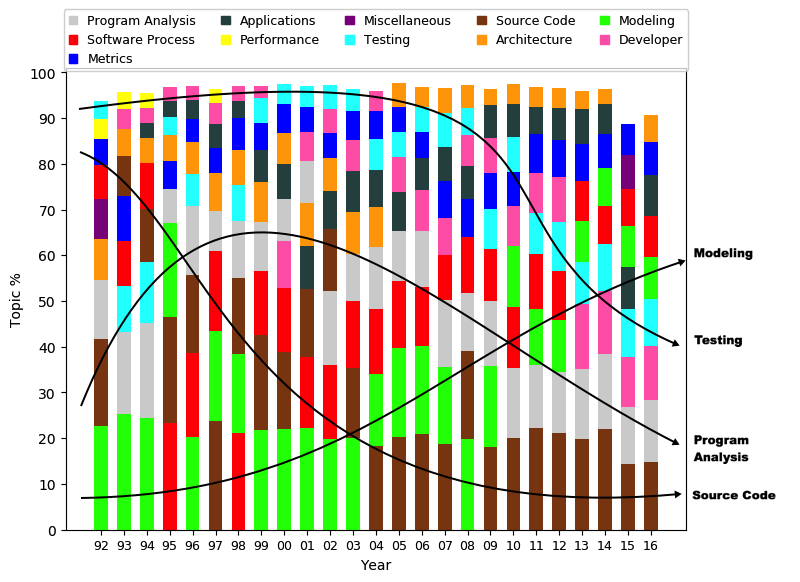}}
\caption{Changes in {\bf conference} topics, 1992-2016.
Arrows are used to highlight the major trends.
This is a {\em stacked} diagram where, for each year, the {\em most}
popular topic appears at the {\em bottom} of each stack. That is,
as topics become {\em more} frequent, they fall towards the {\em bottom} of the plot. Conversely, as topics become {\em less} frequent,
then rise towards the {\em top}.}\label{fig:topic_evo_conferences}

{\centering \includegraphics[scale=0.45]{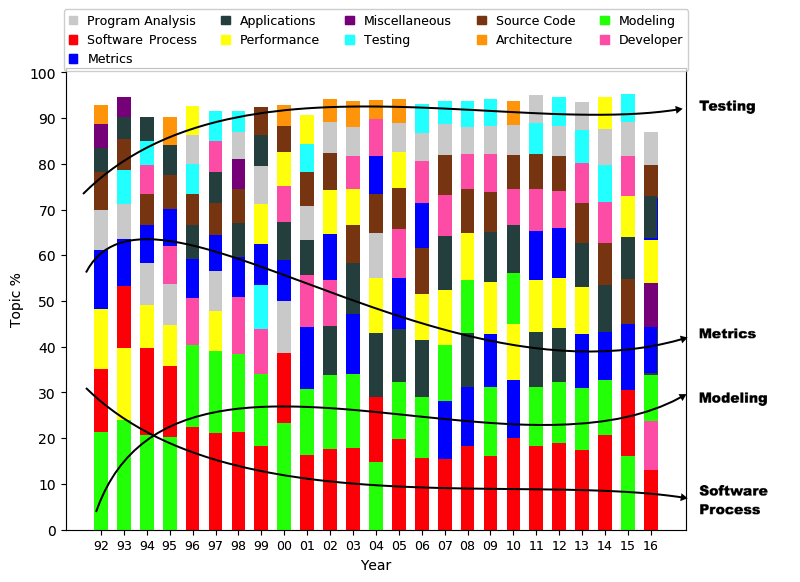}}
\caption{Changes in {\bf journal} topics, 1992-2016.
Same format as \fig{topic_evo_conferences}.
Note that some popular  conference topics (e.g. source code, program analysis, testing) are currently {\em not}
popular  journal topics. }\label{fig:topic_evo_journals}
\end{figure}

\fref{topic_evo_conferences} and \fref{topic_evo_journals} show how the topics change  in conferences and journals over the 
last 25 years of software engineering.
These are {\em stacked diagrams} where, for each year, the {\em more popular} topic appears 
{\em nearer the bottom} of each stack. 
Arrows are added to highlight major trends in our corpus:
\bi
\item {\em Downward} arrows denote topics of {\em increasing} popularity.;
\item  {\em Upward} arrows show topics of {\em decreasing} popularity.
\ei
One clear trend in both figures is that some topics are far
more popular than others.
For example, at the conferences (\fref{topic_evo_conferences}):
\bi
\item
Source code, program analysis, and testing  have become very popular topics in the last decade. 
\item
Performance and security concerns   have all  disappeared in our sampled venues. Performance appears, a little, in journals but is far less popular that many other topics. 
\item Modeling is a topic occurring with decreasing frequency in our sample.
\ei
This is {\em not} to say that performance,  modeling  and security research   has ``failed'' or that no one works on this topics anymore.
Rather, it means that those communities have  moved out of mainstream SE to establish their own niche communities. 

As to the low occurrence of performance and the absence of any security terms in \tref{11topics}, this is one aspect of these results that concerns us. While we hear much talk about security at the venues listed in \tref{venues}, 
\fref{topic_evo_conferences} and \fref{topic_evo_journals}, 
security research is not a popular topic in mainstream software engineering.
Given the current international dependence on software, and all the security violations reported due to software errors\footnote{See \url{https://catless.ncl.ac.uk/Risks/search?query=software+security}.}, 
it is surprising and somewhat alarming  that security is not more prominent in our community. This is a clear area where we would recommend rapid action;  e.g.,
\bi
\item Editors of software journals might consider: increasing the number of special issues devoted to software security;
\item Organizers of SE conferences might consider changing the focus of their upcoming events.
\ei
\subsection{Breadth vs Depth?}
\label{sect:study_jack}

What patterns of research
are most rewarded in our field?
For greater recognition, should a researcher focus on multiple topics or stay dedicated to just a few? We use \fref{sa_top} to answer this question. This figure
compares all authors to those with 1\%, 10,\%, and 20\% of the top ranked authors based on their total number of citations. The x-axis represents the cumulative number of topics covered (ranging from 1-10). The y-axis shows the number of authors working on $x_i$ topics where $x_i$ is a value on the x-axis.  The patterns in the figure are quite clear

\bi
\item More authors from the corpus focus on fewer number of topics.
\item This trend reverses when it comes to the top authors. We can see that in the top 1\% and 10\% of authors (and to a lesser degree in the top 20\%), more authors  focus on fewer number of topics.
\item If an author focuses on all the topics, almost always the author would be in the top 20\% of the SE research community.
\item Very few authors (3 to be precise) in the top 1\% authors focus specifically on a single topic.
\ei

From this, we offer the following advice. The field of SE
is very broad and constantly
changing.
Some flexibility in research
goals tend  to be most rewarding for researchers who explore multiple research directions.
\begin{figure}
\centering
\includegraphics[scale=0.5]{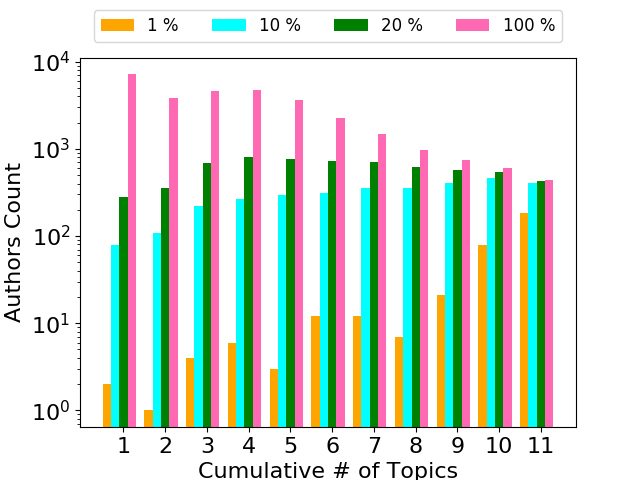}
\caption{Represents number of authors (on log scale) exploring different number of topics for the top 20\%, 10\%, and 1\% authors based on citation counts.}\label{fig:sa_top}
\end{figure}

\subsection{Gender Bias?}
\label{sect:study_bias}

Recent research has raised the spectre
of gender bias in research publications~\cite{Terrell2017GenderDA,roberts2016double,west2012women}. In our field, this is a major concern. For example,
a  recent study by Roberts \etal show that conference reviewing methods have marked an
impact on the percentage of women with
accepted SE papers~\cite{roberts2016double}. Specifically, they showed that the acceptance of papers coauthored by women increased after the adoption of the double blind technique.

Accordingly, this section checks if our
data can comment on gender bias in SE research. The conjecture of this section
is that if the percentage of women authors is $W_0\%$, then the percentage of women in top publications
should also be  $W_1 \approx W_0\%$. Otherwise,
that would be evidence that while this field is
accepting to women authors, our field is also
denying those women senior status in our
community.

Based on our survey of 35,406 authors in our corpus, the ratio of women in SE was increasing till 2004 and since then it has stayed around 22\%. This can be inferred from \fref{gender_per_year} which is a bar chart representing percentage of women in SE each year. Thus, for this study we can say that $W_0 = 22\%$

\fref{genderDiffInfl} shows a line chart with the percentage of women $W_1$ in the top 10 to 1000
most-cited papers in SE (red solid line). The chart also shows the expected value of the percentage of women $W_0$ (blue dashed line). These figures gives us mixed results
\bi
\item On the upside 23\% of the authors in the top 100+ authors are women.
\item But on the downside only 15\% authors in the top 20 are women and it drops down to 10\% in the top 10.
\ei

Although there is an evident bias towards the male community when it comes to the upper echelon of SE, on a larger scale there is evidence of increasing participation in SE research. This is observed in the rise of female authors in SE from less than 8\% in 1991 to close to 22\% in 2015 and over 23\% of the top 100+ authors in SE being female. When compared to other traditional areas of research, the SE field is much ahead (mathematics - 11.2\% and economics - 13.1\%). But there is no denying that the SE community is well behind other scientific fields like veterinary medicine(35.1\%) and cognitive science (38.3\%) where more women are collaborating in research~\cite{west2012women}.

This section has explored
one  specific bias within the SE
community.
Apart from the above,
there may well be other kinds of
biases 
 that systematically
and inappropriately disadvantage
stratifications within our community.
One role for the tools discussed in this paper allows us to 
introspect and detect those
other kinds of biases.

\begin{figure}[!t]
\centering
\includegraphics[scale=0.45]{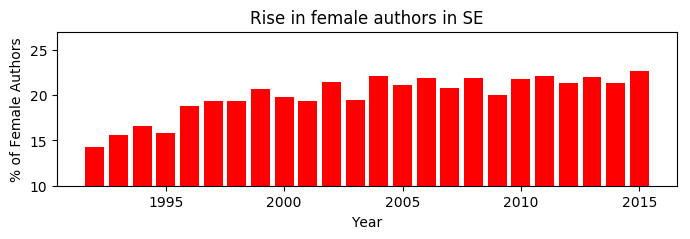}
\caption{\% of female authors from 1991-2015. From the figure, we can see that the SE community saw a great rise in female authors until the late 90s. Since then, the percentage of women in SE has remained around 20-23\%}
\label{fig:gender_per_year}
\end{figure}

\begin{figure}[!t]
\centering
\includegraphics[scale=0.5]{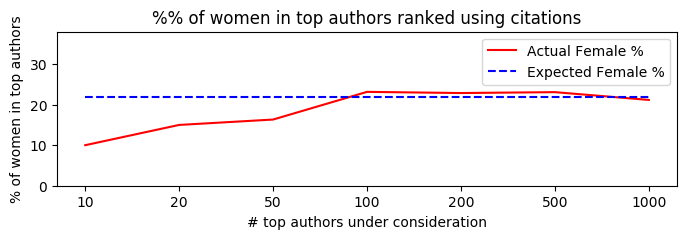}
\caption{\% of women in top 10 to 1000 authors in SE using the total number of citations. Red solid line represents the actual value ($W_1$). While the blue dotted line represents the expected value ($W_0$) which is the percentage of women in SE.}
\label{fig:genderDiffInfl}
\end{figure}

\subsection{Where to publish: journals or conference?}
\label{sect:study_conf_vs_jour}

\fref{topic_evo_conferences} and \fref{topic_evo_journals}   report sharp differences between the kinds of topics published at SE journals and conferences.
For example:
\bi
\item In journals,   metrics, modeling and {software process} appear quite often. But in conferences they appear either rarely or (in the case of modeling) remarkably decreasing frequency in the last decade.
\item In conferences, source code, program analysis, and testing appear to have become very prominent in the last decade but these topics barely appear in journals.
\ei
Hence, we say that SE journals and SE conferences clearly accept different kinds of papers. 
Is this a problem?  Perhaps so. Using our data, we can compute the   number of citations per year for papers published at different venues.
Consider \tref{venue_vs_cites} which shows the median (\textbf{Med}) and inter-quartile range (\textbf{IQR}) for the average number of citations per year for conferences and journals.
\tref{venue_vs_cites} shows that the median of average cites per year 
(in the future for the papers published in a given year) has been steadily increasing since 2000.
When same data from \tref{venue_vs_cites} is looked at but this time broken down into conferences and journals:
\bi
\item A \colorbox{rc1}{red} background indicates when conferences are  receiving  more citations per year;
\item A \colorbox{rc2}{green}  background indicates when journals papers are receiving more citations per year;
\item Otherwise, there is no statistically difference between the two distributions.
\ei
For this analysis, ``more citations'' means that the distributions are statistically significantly different (as judged via  the 95\% confident bootstrap procedure
recommended by   Efron \& Tibshirani~\cite[p220-223]{efron93}) and that difference
is not trivially small (as judged by the A12 test endorsed by Arcuri et al. at ICSE '11 \cite{arcuri11}).

 \begin{table} 
\scriptsize
\centering
\begin{tabular}{|l|cc|cc|}
\hline
\multicolumn{1}{|c|}{\multirow{2}{*}{\textbf{Year}}} & \multicolumn{2}{c|}{\textbf{Conference}} &
\multicolumn{2}{c|}{\textbf{Journal}} \\
 \multicolumn{1}{|c|}{} & \textbf{Med} & \textbf{IQR} & \textbf{Median} & \textbf{IQR}  \\ \hline
 \rB 1992 & 0.08 & 0.20 & 0.12 & 0.60 \\ 
 \rB 1993 & 0.08 & 0.33 & 0.13 & 0.46 \\
 \rB 1994 & 0.09 & 0.35 & 0.13 & 0.43 \\
     1995 & 0.27 & 0.64 & 0.14 & 0.45 \\
     1996 & 0.10 & 0.33 & 0.05 & 0.48 \\
     1997 & 0.10 & 0.30 & 0.10 & 0.50 \\
     1998 & 0.26 & 0.63 & 0.11 & 0.63 \\
     1999 & 0.06 & 0.28 & 0.06 & 0.50 \\
     2000 & 0.12 & 0.59 & 0.06 & 0.47 \\
 \rB 2001 & 0.13 & 0.50 & 0.25 & 1.00 \\
 \rB 2002 & 0.20 & 0.60 & 0.27 & 1.13 \\
     2003 & 0.29 & 0.71 & 0.36 & 1.14 \\
 \rB 2004 & 0.23 & 0.69 & 0.38 & 1.23 \\
     2005 & 0.33 & 0.75 & 0.25 & 0.92 \\
 \rB 2006 & 0.18 & 0.82 & 0.45 & 1.36 \\
 \rB 2007 & 0.40 & 1.00 & 0.50 & 1.50 \\
 \rB 2008 & 0.44 & 1.11 & 0.78 & 1.78 \\
 \rB 2009 & 0.38 & 1.00 & 0.63 & 1.75 \\
     2010 & 0.57 & 1.43 & 0.71 & 1.71 \\
     2011 & 0.67 & 1.50 & 0.67 & 1.83 \\
 \rA 2012 & 0.80 & 2.00 & 0.80 & 1.60 \\
 \rA 2013 & 0.75 & 2.50 & 0.75 & 1.75 \\
 \rA 2014 & 1.00 & 2.00 & 0.67 & 2.00 \\
 \rA 2015 & 1.00 & 2.00 & 0.50 & 2.00 \\ 
 \hline
\end{tabular}
\\
\begin{tabular}{cc}
\\
\legend{rc1}{Conference} & \legend{rc2}{Journal}
\end{tabular}

\caption{Median (\textbf{Med}) and Inter Quartile Range (\textbf{IQR}=inter-quartile range= (75-25)th range) of average cites per year for articles published in conferences and journals between 1992-2015. Column one colors denote the years when either   \colorbox{rc1}{Conferences} or \colorbox{rc2}{Journals} 
received a statistically significantly larger number of citations per year.  }
\label{tab:venue_vs_cites}
\end{table}

 
Note that until very recently (2009), journal papers always achieved statistically
more cites per year than conference papers. However, since 2012, conference papers now receive more cites per year. These statistics justify Fernandes' 2014 study~\cite{fernandes2014authorship} where he states that since 2008, there has been a significant increase (almost double) in the number of conference publications over journal articles. 

 There any many actions journal editors could undertake to mitigate this trend. For example, a recent trend in SE conferences are the presentation of ``Journal-First'' papers 
 that have already appeared in journals. Since they are being seen by a larger audience, such journal-first papers might receive more citations per year. Also, once a journal
 becomes known for a Journal-first program, then that could increase the number of submissions to that journal.

\section{Threats To Validity}
\label{sect:threats}

This paper is less prone to {\em tuning bias} and {\em order bias} that other software analytics papers. As discussed earlier in section \ref{sect:lda}, we use differential evolution o find our tunings.

The main threat to validity of our results is {\em sampling bias}. The study can have two kinds of sampling bias
\bi
    \item This study only accessed the accepted papers
{\em but not the rejected ones}. Perhaps if both accepted and rejected papers
are studied, then some patterns other that the ones reported here might be discovered. That said, sampling those rejected papers is problematic. We can access 100\% of accepted
papers via on-line means. Sampling rejected papers imply asking researchers to supply their rejected
papers. The experience in the research community with such surveys is that only a small percent of respondents reply~\cite{robles2014estimating}
in which case we would have another sampling bias amongst the population of rejected
papers. Additionally, most researchers alter their rejected papers in-line with an alternate conference/journal and make a new submission to the venue. At the time of this writing, we do not know how to resolve this issue.
\item The paper uses 34 top venues considering their online citation indices and feedback from the software engineering community (see Section \ref{sect:data}). Thus, there will always be a  venue missing from such a study and can be considered. This can raise questions on the sanity of the 10 topics which model SE. But, the scale of a study to consider all the publications even remotely related to SE is extremely large. Hence in section \ref{sect:topicSanity} we show that the topics remain similar when a reduced set of venues are used. Thus, this bias can be overcome to a great extent if a diverse set of venues are considered.
\ei
{Section \ref{sect:study_bias} discussed if there exists a  gender bias in SE research. There, we commented
that, measured in terms of women participation,  the SE community seems to be falling behind fields like veterinary medicine and cognitive science. This might be due to the fact there could be a lower influx of women into graduate studies, thereby producing a lower pool of potential researchers. An alternative explanation might be
due to   students stopping their studies before graduation due to the highly rewarding industrial career in Computer Science for undergraduates (compared to other fields). More research is needed on this point.}

\section{Related Work}
\label{sect:rel_work}

To the best of our knowledge, this paper is the largest study of the SE literature yet accomplished (where ``large'' is measured in terms of number of papers and number of years). 

Unlike prior work~\cite{glass2005assessment,ren2007automatic,wohlin2007analysis,garousi2013bibliometric}
our analysis is fully automatic and repeatable. The same is not true for other studies.
For example, the Ren and Taylor~\cite{ren2007automatic} method of ranking scholars and institutions incorporate manual weights assigned by the authors. Those weights were dependent on expert knowledge and has to be updated every year.  
 Our automation means that this analysis is quickly
 repeatable whenever new papers are published.

\subsection{Bibliometrics Based Studies}
\label{sect:biblio}

Multiple bibliometric based studies have explored patterns in SE venues over the last decades. Some authors have explored only conferences~\cite{systa2012inbreeding, vasilescu2014healthy} or journals~\cite{hamadicharef2012scientometric} independently while some authors have explored a mixture of both~\cite{ren2007automatic, cai2008analysis, fernandes2014authorship}.

Early bibliometric studies were performed by Ren \& Taylor  in 2007  where they assess both academic and industrial research institutions, along with their scholars to identify the top ranking organizations and individuals~\cite{ren2007automatic}. They provide an automatic and versatile framework using electronic bibliographic data to support such rankings which produces comparable results as those from manual processes. This method although saves labor for evaluators and allow for more flexible policy choices, the method does not provide a spectrum of the topics and the publication trends in SE.

Later in 2008, Cai \& Card~\cite{cai2008analysis} analyze 691 papers from 14 leading journals and conferences in SE. They observe that 89\% of papers in SE focus on 20\% of the subjects in SE, including software/program verification, testing and debugging, and design tools and techniques. We repeat this study over a wider spread of publications, venues and authors (see \tion{data})  and the results conform with their observations.

In 2012, Hamadicharef performed a scientometrics study on IEEE Transactions of Software Engineering (TSE) for three decades between 1980-2010~\cite{hamadicharef2012scientometric}. He analyzes five different questions
\bi
\item Number of publications: A quasi-linear growth in number of publications each year.
\item Authorship Trends: It is observed that, a large number of articles in the 80s had only one or two co-authors but since the turn of the century it seemed to increase more papers having 5, 6 or 7 co-authors.
\item Citations: The most cited TSE paper in this period had 811 cites and on an average a paper was cited 22.22 times with a median of 9 cites. On the other hand 13.38\% papers were never cited. A larger study over 34 venues was repeated in this paper comparing the citation trends between conferences and journals (See \tion{study_conf_vs_jour}).
\item Geographic Trends: 46 different countries contribute to TSE between this period. 57\% of the contributions to TSE come from USA, close to 21\% come from Europe and less than 1\% publications come from China.
\item \# of References: The average number of references per article increases 12.9 to 42.9 between 1980 to 2010.
\ei
A much larger study using 70,000 articles in SE was conducted in 2014 by Fernandes~\cite{fernandes2014authorship}. He observes that the number of new articles in SE doubles on an average every decade and since 2008 conferences publish almost twice as many papers as journals every year. Though the paper fails to address how the citation trends have been varying between conferences and journals and if it attributes towards the increased number of publications in conferences over journals. 

More recently, Garousi and Fernandes~\cite{garousi2016highly} \etal  performed a study based on citations to identify the top cited paper in SE. This study was based on two metrics: a) total number of citations and b) average annual number of citations to identify the top papers. The authors also go to the extent of characterizing the overall citation landscape in Software Engineering hoping that this method will encourage further discussions in the SE community towards further analysis and formal characterization of the highly-cited SE papers. 

Geographical based bibiliometric studies on Turkish~\cite{garousi2015bibliometric} and Canadian~\cite{garousi2010bibliometric} SE communities were performed by Garousi \etal to study the citation landscape in the respective countries. They identify a lack of diversity in the general SE spectrum, for example, limited focus on requirements engineering, software maintenance and evolution, and architecture. They also identify a low involvement from the industry in SE. Since these studies were localized to a certain country, it explored lesser number of papers

Citation based studies have also evolved into adoption of measures such as h-index and g-index to study the success of an SE researcher. Hummel \etal in 2013 analyzed the expressiveness of modern citation analysis approaches like h-index and g-index by analyzing the work of almost 700 researchers in SE~\cite{hummel2013analyzing}. They concluded that on an average h-index for a top author is around 60 and g-index is about 130. The authors used citations to rank authors and some researchers are apprehensive to this definition of success~\cite{fernandes2014authorship, ding2009pagerank}.

Vasilescu \etal studied the health of SE conferences with respect to community stability, openness to new authors, inbreeding, representatives of the PC with respect to the authors’ community, availability of PC candidates and scientific prestige~\cite{vasilescu2014healthy}. They analyzed conference health using the R project for statistical computing to visualize and statistically analyze the data to detect patterns and trends~\cite{team2013r}. They make numerous observations in this study
\bi
    \item Wide-scoped conferences receive more submissions, smaller PCs and higher review load compared to narrow scoped conferences.
    \item Conferences considered in the study are dynamic and have greater author turnover compared to its previous edition.
    \item Conferences like ASE, FASE and GPCE are very open to new authors while conferences like ICSE are becoming increasingly less open.
    \item Lesser the PC turnover, greater the proportion of papers accepted among PC papers.
    \item Narrow-scoped conferences have more representative PCs than wide-scoped ones.
    \item Not surprisingly, the higher the scientific impact of a conference, the more submissions it attracts and tend to have lower acceptance rates. The authors work is very detailed and gives a detailed summary of SE conferences.
\ei
Although this paper is very extensive, the authors do not explain the topics associated with venues (\tion{topicLabel} and \tion{study_hot})  or how conferences and journals are similar/different from each other(\tion{study_conf_vs_jour}).

\subsection{Topic Modeling}

Another class of related work are {\em Topic Modeling} based studies which has been used in various spheres of Software Engineering. According to a survey reported by Sun \etal~\cite{sun2016exploring}, topic modeling is applied in various SE tasks, including source code comprehension, feature location, software defects prediction, developer recommendation, traceability link recovery, re-factoring, software history comprehension, software testing and social software engineering. There are works in requirements engineering where it was necessary to analyze the text and come up with the important topics~\cite{asuncion2010software, thomas2014studying, massey2013automated}. People have used topic modeling in prioritizing test cases, and identifying the importance of test cases~\cite{hemmati2015prioritizing, zhang2015inferring, yang2015predicting}. Increasingly, it has also become very important to have automated tools to do SLR~\cite{tsafnat2014systematic}. We found these papers~\cite{restificar2012inferring,alreview,marshall2013tools} who have used clustering algorithms (topic modeling) to do SLR.

Outside of SE, in the general computer science (CS) literature, a 2013 paper by Hoonlor \etal highlighted the prominent trends in CS~\cite{hoonlor2013trends}. This paper identified trends, bursty topics, and interesting inter-relationships between the American National Science Foundation (NSF) awards and CS publications, finding, for example, that if an uncommonly high frequency of a specific topic is observed in publications, the funding for this topic is usually increased. The authors adopted a Term Frequency Inverse Document Frequency (TFIDF) based approach to identify trends and topics. A similar approach can be performed in SE considering how closely CS is related to SE.

Garousi and Mantyla recently have adopted a Topic Modeling and Word Clustering based approach to identify topics and trends in SE~\cite{garousi2016citations} similar to the research by Hoonloor \etal\cite{hoonlor2013trends}. Although their method is very novel and in line with the current state of the art, but they used only the titles of the papers for modeling topics. This might lead to inaccurate topics as titles are generally not very descriptive of the field the paper is trying to explore. This issue is addressed in the current work where we use the abstracts of the paper which gives more context while building models.

In 2016, Datta \etal\cite{datta2016long} used Latent Dirichlet Allocation to model 19000 papers from 15 SE publication venues over 35 years into 80 topics and study the half life of these topics. They coin the term ``Relative Half Life" which is defined as the period between which the ``importance" of the topic reduces to half. They further define two measures of importance based on the citation half life and publication half life. The choice of 80 topics is based on lowest log likelihood and although very novel but the authors do not shed light on the individual topic and the terms associated with it. Note that we do not
recommend applying their kind of analysis since it lacks
automatic methods for selecting the control parameters for LDA whereas our method is automatic.

\section{Conclusions}
\label{sect:conclusion}  
Here,     text mining methods were applied to 35,391 documents written in the last 25 years from 34 top-ranked SE venues. These venues were divided, nearly evenly, between conferences and journals.
An important aspect of this analysis is that it is fully automated. Such automation allows
for the rapid confirmation and updates of these results, whenever
new data comes to hand.
To achieve that automation, we used a   topic modeling technique called LDA (augmented with an automatic 
assistant for tuning the control parameters called differential evolution).

Sometimes we are asked what is the point of work like this?  ``Researchers,'' say some, ``should be free to explore
whatever issues they like, without interference from some burdensome supervision body telling them
what they should, or should not conduct particular kinds of research''.
While this is a valid point,  we would say that 
this actually  endorses the need for the research in this paper. 
We fully accept and  strongly endorse the principle that researchers should be able to explore
software engineering, however their whims guide them. But if the    structure of SE venues is inhibiting, then that structure should change. This is an important point since, as discussed above,
there are some troubling patterns within the SE literature:
\bi
\item There exists  different sets of
topics that tend to be accepted to SE conferences or journals;
researchers exploring some topics contribute more towards certain venues.
\item
We show in \tion{study_conf_vs_jour} that a recent trend where SE conference papers are
receiving significantly larger citations per year than journal
papers. For academics whose career progress gets reviewed (e.g., during the tenure process; or if ever those academics are applying for new jobs),
it is important to know what kinds of venues adversely affect citation counts.
\item
We also highlighted gender bias issues that need to be explored more in future work.
\ei
We further suggest that, this method can also be used for strategic and tactical purposes.
    Organizers of SE venues could   use them as a long-term planning aid
    for improving and rationalizing how they service 
    our research community. Also, individual researchers could also use these results
    to make short-term publication plans. 
    Note that our analysis is 100\% automatic,
    thus making it readily repeatable and easily be updated.

It must further be stressed
that, the   topics reported here
should not be ``set in stone''
as the only  topics   worthy of study in SE.
Rather our goal is to report that   (a)~text mining methods can  detect large scale trends within our community;  (b)~those topics
 change with time; so (c)~it is  important to have automatic agents that can update our understanding
of our community
whenever new data arrives.

\bibliographystyle{IEEEtranS}

\bibliography{IEEEabrv,refs}

\newpage

 
\begin{IEEEbiography}[{\includegraphics[width=1in,clip,keepaspectratio]{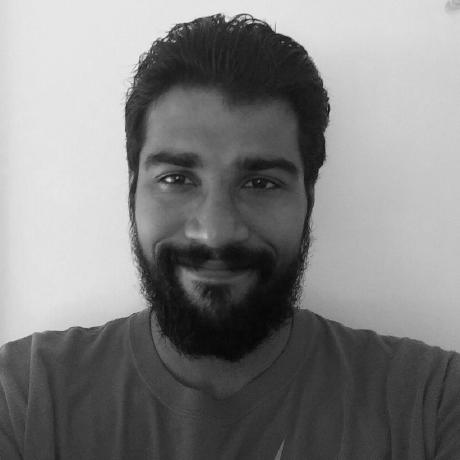}}]{George Mathew}
is a third year PhD Student of the Department of Computer Science at North Carolina State University.
With industrial experience at Facebook, CrowdChat and Microsoft, Mr. Mathew
explores automated program repair, multi-objective optimization and 
applied machine learning algorithms in software engineering. For more information on his research and interests, see \url{http://bigfatnoob.us}
\end{IEEEbiography}

\begin{IEEEbiography}[{\includegraphics[width=1in,keepaspectratio]{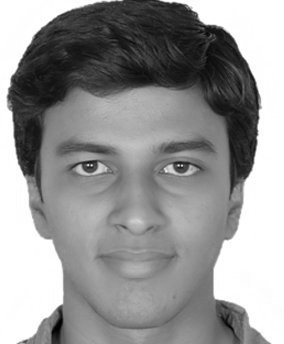}}]{Amritanshu Agrawal}
is a fourth year PhD Student of the Department of Computer Science at North Carolina State University, Raleigh, NC. 
His research involves defining better, and simpler, quality improvement operators for software analytics. 
He aims at providing better next generation AI software to software engineers by making them literate with AI. He has industrial experiences at IBM, Lucidworks and LexisNexis. For more, see
\url{http://www.amritanshu.us}.
\end{IEEEbiography}

\begin{IEEEbiography}[{\includegraphics[width=1in, clip]{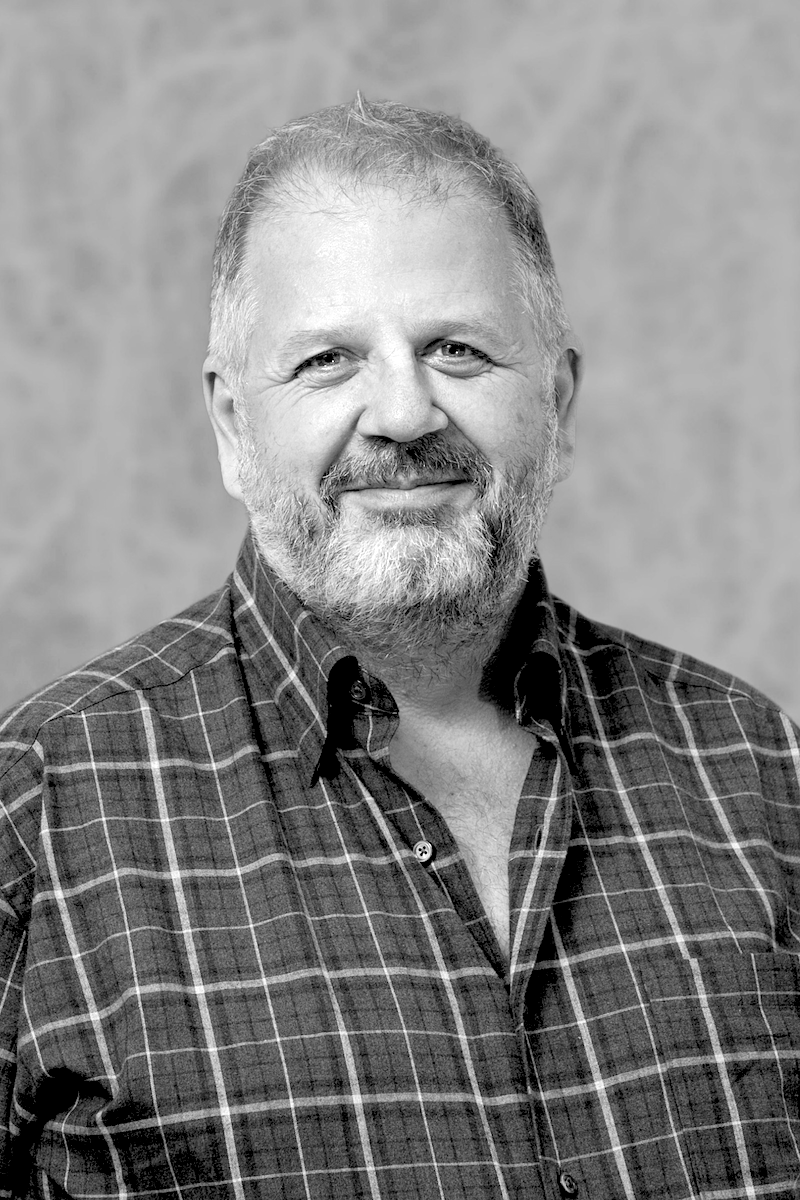}}]
{Tim Menzies} (Ph.D., UNSW, 1995) is a full Professor in CS at NC State University, where he explores SE, data mining, AI, search-based SE, programming languages and open access science. He is the author of over 250 referred publications and was co-founder of the ROSE festivals and PROMISE conferences devoted to reproducible experiments in SE (\url{http://tiny.cc/seacraft}).  Dr. Menzies does, or has,  served, as associated editor of many journals:  IEEE TSE, ACM TOSEM, Empirical Software Engineering, ASE Journal, Information Software Technology, IEEE Software, and the Software Quality Journal. For more, see \url{http://menzies.us}.
\end{IEEEbiography}
\balance


\end{document}